\documentclass[amsfonts,showpacs,tightenlines,aps,12pt,floatfix]{revtex4}
\usepackage{bm}
\usepackage{epsfig}
\usepackage{graphicx}
\begin{document}


\title{The Realistic Lattice Determination of $\alpha_s(M_Z)$ Revisited}
\author{K. Maltman}
\email[]{kmaltman@yorku.ca}
\affiliation{Department of Mathematics and Statistics, York University, 
4700 Keele St., Toronto, ON CANADA M3J 1P3}
\altaffiliation{CSSM, School of Chemistry and Physics,
Univ. of Adelaide, SA 5005 AUSTRALIA}
\author{D. Leinweber, P. Moran and A. Sternbeck}
\email[]{dleinweb@physics.adelaide.edu.au,andre.sternbeck@adelaide.edu.au,
peter.moran@adelaide.edu.au}
\affiliation{CSSM, School of Chemistry and Physics,
University of Adelaide, SA 5005 AUSTRALIA}

\date{\today}

\begin{abstract}
We revisit the earlier determination of $\alpha_s(M_Z)$ via perturbative
analyses of short-distance-sensitive lattice observables, incorporating
new lattice data and performing a modified version of
the original analysis. We focus on two high-intrinsic-scale observables, 
$\log(W_{11})$ and $\log(W_{12})$, and one lower-intrinsic-scale 
observable, $\log(W_{12}/u_0^6)$, finding improved consistency among 
the values extracted using the different observables and a final result, 
$\alpha_s(M_Z)=0.1192\pm 0.0011$, $\sim 2\sigma$ higher 
than the earlier result, in excellent agreement 
with recent non-lattice determinations and, in addition,
in good agreement with the results of a similar, but not identical, 
re-analysis by the HPQCD collaboration. 
A discussion of the relation between the two re-analyses is given,
focussing on the complementary aspects of the two approaches.

\end{abstract}


\maketitle

\section{\label{intro}Introduction}
The strong coupling $\alpha_s$ is usually characterized by giving the
value, $\alpha_s(M_Z)$, in the $\overline{MS}$ scheme at the conventionally
chosen $n_f=5$ reference scale $\mu =M_Z$. A high precision determination
of $\alpha_s(M_Z)$ based on the perturbative analysis of 
short-distance-sensitive lattice observables computed using
the $a\sim 0.09$, $0.12$ and $0.18$ fm $n_f=2+1$ MILC data was presented in
Ref.~\cite{latticealphas}. The result, $\alpha_s(M_Z)=0.1170(12)$, 
plays a dominant role in fixing
the central value of the current PDG assessment~\cite{pdg06qcdreview},
$\alpha_s(M_Z)=0.1176(20)$.

Over the last year, a number of improved non-lattice determinations of
$\alpha_s(M_Z)$ have appeared, in a variety of independent processes, 
over a wide range of 
scales~\cite{bck08,davieretal08,alphasherajetcrosssections,alphash1jetshighqsq08,alphaseventshapes08,bs08,kst07,brambillaetalupsilon07,jb08,my08}.
The results, given in Table~\ref{table1} (with all errors combined in 
quadrature), yield a weighted average, 
$\alpha_s(M_Z)=0.1190(10)$, $\sim 2\sigma$ higher than the 
lattice determination. This difference, though not large,
motivates revisiting the lattice analysis,
especially in light of the existence of new high-scale ($a\sim 0.06$ fm) 
lattice data not available at the time of the earlier study.
We perform such an extended re-analysis in this paper.

\begin{table}
\caption{\label{table1}Recent non-lattice determinations of
$\alpha_s(M_Z)$}
\vskip .1in
\begin{tabular}{|ll|}
\hline
Source&$\alpha_s(M_Z)$\\
\hline
Global EW fit~\cite{bck08,davieretal08}&$0.1191\pm 0.0027$\\
H1+ZEUS NLO inclusive jets~\cite{alphasherajetcrosssections}&
$0.1198\pm 0.0032$\\
H1 high-$Q^2$ NLO jets~\cite{alphash1jetshighqsq08}&$0.1182\pm 0.0045$\\
NNLO LEP event shapes~\cite{alphaseventshapes08}&$0.1240\pm 0.0033$\\
NNNLL ALEPH+OPAL thrust distributions~\cite{bs08}&$0.1172\pm 0.0022$\\
$\sigma [e^+e^-\rightarrow hadrons]$ (2-10.6 GeV)~\cite{kst07}&
$0.1190^{+0.0090}_{-0.0110}$\\
${\frac{\Gamma [\Upsilon (1s)\rightarrow \gamma X]}
{\Gamma [\Upsilon (1s)\rightarrow X]}}$~\cite{brambillaetalupsilon07}&
$0.1190^{+0.0060}_{-0.0050}$\\
hadronic $\tau$ decay~\cite{jb08,my08,tauentryfootnote}&$0.1187\pm 0.0016$\\
\hline
\end{tabular}
\end{table}

The rest of the paper is organized as follows. In Section~\ref{sec2},
we outline the original analysis, specify our own strategy 
for implementing the underlying approach, and clarify the difference between 
our implementation and that of the earlier study and recent HPQCD 
re-analysis. In Section~\ref{sec3}, we discuss the details of, and 
input to, our version of the analysis. Finally, in Section~\ref{sec4}, 
we present and discuss our results.

\section{\label{sec2}The lattice determination of $\alpha_s(M_Z)$}

\subsection{\label{subsec1}The original HPQCD/UKQCD analysis}

In Ref.~\cite{latticealphas}, $\alpha_s(M_Z)$ was extracted
by studying perturbative expansions for a number of UV-sensitive
lattice observables, $O_k$. The generic form of this expansion is
\begin{equation}
O_k=\sum_{N=1}\bar{c}_N^{(k)}\alpha_V(Q_k)^N
\equiv D_k\alpha_V(Q_k)\sum_{M=0}c_M^{(k)}\alpha_V(Q_k)^M
\label{3loopPT}\end{equation}
where $Q_k=d_k/a$ are the Brodsky-Lepage-Mackenzie (BLM) scales~\cite{blmscale}
for the $O_k$, and $c_0^{(k)}\equiv 1$. The 
coefficients $\bar{c}_{1,2,3}^{(k)}$ (equivalently, $D_k, c_1^{(k)}$, and 
$c_2^{(k)}$) have been computed in 3-loop lattice perturbation 
theory~\cite{mason}, and, with the corresponding $d_k$, tabulated for a number 
of $O_k$ in Refs.~\cite{latticealphas,mason,hpqcd08}. 
In Eq.~(\ref{3loopPT}), $\alpha_V(\mu )$ is a coupling with the same
expansion to $O(\alpha^3_s)$ (with $\alpha_s$ the $\overline{MS}$ coupling)
as the heavy quark potential coupling, $\alpha_V^p$, 
but differing from it, beginning at $O(\alpha_s^4)$, in a way 
that will be specified below. The expansion coefficients
are known to $O(\alpha_s^4)$, and hence the
$\beta$ function of $\alpha_V$, defined in
our conventions by $\mu^2 da_V(\mu )/d\mu^2\, 
=\, -\sum_{n=0}\beta^V_na^{n+2}_V(\mu )$, with $a_V\equiv \alpha_V/\pi$, 
is determined to 4 loops by the known coefficients, 
$\beta_0,\, \cdots,\, \beta_3$, of the
4-loop $\overline{MS}$ $\beta$ function~\cite{4loopbeta}.
The coefficients $\bar{c}^{(k)}_1$, $\bar{c}^{(k)}_2$, and 
$\bar{c}^{(k)}_3$ tabulated in Refs.~\cite{latticealphas,mason,hpqcd08}
are valid for expansions of the $O_k$
in terms of any variable, $\alpha_T$, sharing the same 
expansion as $\alpha_V$ out to $O(\alpha_s^3)$.

With only the known, third order terms in the expansions of
the $O_k$, no value for the reference scale coupling, 
$\alpha_V(7.5\ {\rm GeV})\equiv\alpha_V^0$, was found to
produce a simultaneous fit to the data at all three lattice spacings
employed~\cite{latticealphas}. In consequence, terms out 
to tenth order in the expansion of Eq.~(\ref{3loopPT}) were 
incorporated, the unknown coefficients $\bar{c}_{4,\cdots ,10}^{(k)}$ 
being fitted using input Bayesian prior constraints. The 4-loop 
version of $\beta^V$ was used to run $\alpha_V^0$ to the scales $Q_k$ 
relevant to each of the given observables at each of the three lattice 
spacings. Linear extrapolation in the quark masses was employed, and
possible residual mass-independent non-perturbative (NP) contributions 
estimated, and subtracted, using the known leading-order gluon condensate
contributions to the relevant Wilson loops~\cite{wmnglue}.

The scales $r_1/a$ and $r_1$, which determine the lattice spacing,
$a$, in physical units, as well as the gluon condensate,
$\langle \alpha_s G^2/\pi \rangle$, 
required for the mass-independent NP subtraction, were
determined as part of the independent fit performed for each of the
$O_k$. This was accomplished using 
an augmented $\chi^2$ function in which the squared deviations of
the relevant parameters from their input central values were
scaled by the squares of the input prior widths. For $r_1/a$ and
$r_1$ the central values and widths were provided by the measured values and
their uncertainties. For $\langle \alpha_s G^2/\pi \rangle$, a central value
$0$ and uncertainty $\pm 0.010\ {\rm GeV}^4$ 
($\sim$ the conventional SVZ value $0.012\ {\rm GeV}^4$~\cite{svz}) 
were employed~\cite{aggthanks}. While this procedure allows 
$r_1/a$ and $r_1$ (which should be characteristic of the lattice 
under consideration) to take on values which vary slightly 
with the $O_k$ being analyzed, one should bear in mind that 
the measured uncertainties, which set the range of these variations, 
are small compared to the variation of scales across the 
$a\sim 0.09, 0.12$ and $0.18$ fm lattices employed in the analysis.
The impact of any potential unphysical
observable-dependence of the physical scales on the fitted 
$\alpha_V^0$ and $\bar{c}_n^{(k)}$ should thus be safely negligible.
The situation with regard to the independent fitting of
$\langle \alpha_s G^2/\pi \rangle$ for each $O_k$ is potentially
more complicated, and will be discussed further below.

The resulting best fit value for $\alpha_V^0$, averaged over
the various observables, 
was then matched to the $n_f=3$ $\overline{MS}$ coupling,
and the corresponding $n_f=5$ result, $\alpha_s(M_Z)$, obtained via
standard running and matching at the flavor 
thresholds~\cite{cks97,matchingrunningfootnote},
yielding the result, $\alpha_s(M_Z)=0.1170(12)$, already quoted above. 

Regarding the conversion from $\alpha_V$ to $\alpha_s$,
one should bear in mind that, while the expansion for $\alpha_V$
in terms of $\alpha_s$ is, in principle, defined to all orders
(see below for more on this point), the 
coefficients beyond $O(\alpha_s^4)$ involve 
the currently unknown $\overline{MS}$ $\beta$ function coefficients 
$\beta_4,\, \beta_5,\, \cdots$. 
The $n_f=3$ conversion step is thus subject to a (hopefully small) 
higher order perturbative uncertainty. 
As will be explained in Section~\ref{subsec3}, with the
definition of $\alpha_V$ employed in Ref.~\cite{latticealphas},
the higher order perturbative uncertainties are, in fact,
entirely isolated in the $V\rightarrow\overline{MS}$ 
conversion step of the analysis. 

\subsection{\label{subsec2}An alternate implementation of the
HPQCD/UKQCD approach}

The higher order perturbative uncertainty encountered in 
matching $\alpha_V$ to $\alpha_s$ can be removed entirely by working with any
expansion parameter, $\alpha_T$, whose 
expansion in $\alpha_s$ is fully specified. We
take $\alpha_T$ to be defined by the third-order-truncated
form of the relation between $\alpha_V^p(\mu^2)$ and 
$\alpha_s(\mu^2)$~\cite{schroder} which, for $n_f=3$, yields
\begin{equation}
\alpha_T(\mu^2 ) = \alpha_s(\mu^2 )\left[ 1+0.5570\alpha_s(\mu^2 )
+1.702\alpha^2_s(\mu^2 )\right]\ .
\label{alphatdefn}\end{equation}
The $\beta$ function for $\alpha_T$, $\beta^T$, is then
determined to 4-loops by the known values of $\beta_0,\, \cdots,\,
\beta_3$. With all coefficients on the RHS positive, $\alpha_T$ runs much 
faster than $\alpha_s$, a fact reflected in the significantly
larger values of the non-universal $\beta$ function coefficients, 
$\beta^T_2=33.969$ and $\beta^T_3\, =\, -324.393$. This makes
running $\alpha_T$ using the 4-loop-truncated $\beta^T$ function 
typically unreliable at the BLM scales corresponding to the 
coarsest ($a\sim 0.18$ fm) lattices considered here. 
Since, however, the 4-loop-truncated $\overline{MS}$ running of
$\alpha_s$ remains reliable down to these scales, and
the relation, Eq.~(\ref{alphatdefn}) is, by definition, exact,  
the running of $\alpha_T$ may be performed by converting from 
$\alpha_T$ to $\alpha_s$ at the initial scale, running $\alpha_s$ to the final
scale, and then converting back to $\alpha_T$. This proceedure
will be especially reliable for $)_k$ like $\log (W_{11})$ 
and $\log (W_{12})$ with lowest BLM scales $>3$ GeV.

Though the conversion from the fitted reference 
scale $\alpha_T$ value to the equivalent $\overline{MS}$ coupling $\alpha_s$
can be accomplished without perturbative uncertainties,
higher order perturbative uncertainties do remain in
the analysis. To see where, define
$\alpha_0\equiv \alpha_T(Q_0)$, with $Q_0=Q_k^{max}=d_k/a_{min}$
the maximum of the BLM scales (corresponding to the finest
of the lattice spacings, $a_{min}$) for the observable in question.
Expanding the couplings at those BLM scales corresponding to coarser 
lattices, but the same observable, in the standard manner as a power 
series in $\alpha_0$, $\alpha_T(Q_k)=\sum_{N=1}p_N(t_k)\alpha_0^N$
(where $t_k=\log\left( Q_k^2/Q_0^2\right)$, and the $p_N(t)$ are polynomials 
in $t$), one finds, on substitution into Eq.~(\ref{3loopPT}), 
\begin{eqnarray}
{\frac{O_k}{D_k}}\, &=&\, \cdots +
\alpha_0^4\left( c_3^{(k)}+ \cdots\right) +\alpha_0^5
\left( c_4^{(k)}-2.87 c_3^{(k)} t_k+\cdots\right)
+\alpha_0^6\left( c_5^{(k)} -0.0033 \beta_4^T t_k \right.\nonumber\\
&&\left. \qquad
-3.58 c_4^{(k)} t_k +[5.13 t_k^2 - 1.62 t_k] c_3^{(k)}+\cdots\right)
+\alpha_0^7 \left( c_6^{(k)}
-0.0010 \beta_5^T t_k \right.\nonumber\\ 
&&\left. \qquad +[0.0094 t_k^2-0.0065 c_1^{(k)} t_k] \beta_4^T
-4.30 c_5^{(k)}t_k +[7.69 t_k^2 - 2.03 t_k] c_4^{(k)} 
\right.\nonumber\\
&&\left. \qquad 
+[- 7.35 t_k^3+ 6.39 t_k^2 -4.38 t_k] c_3^{(k)} +\cdots\right)+\cdots\ .
\label{cfitproblem}\end{eqnarray}
where the known numerical values of $\beta_0^T,\cdots ,\beta_3^T$
have been employed, and we display only terms
involving one or more of the unknown quantities $\beta_4^T,\beta_5^T,\cdots$,
$c_3^{(k)},c_4^{(k)},\cdots$.

Running the $\overline{MS}$ coupling numerically using the
4-loop-truncated $\beta$ function is equivalent to keeping
terms involving $\beta_0,\cdots ,\beta_3$ to all orders,
and setting $\beta_4=\beta_5=\cdots =0$. The neglect of
$\beta_4,\beta_5\cdots$ means that $\beta^T_4,\beta^T_5,\cdots$
also do not take on their correct physical values, leading
to an alteration of the true $t_k$-dependence, beginning at $O(\alpha_0^6)$.
Since it is the scale-dependence of $O_k$ which is used to fit
the unknown coefficients $c_{3,4,\cdots}^{(k)}$, as well as
$\alpha_0$, we see immediately that the 4-loop truncation necessarily
forces compensating changes in at least the coefficients 
$c^{(k)}_{4,5,\cdots}$. A shift in $c_4^{(k)}$, however, also alters
the $O(\alpha_0^5)$ coefficient, which will, in general, necessitate an
approximate compensating shift in $c_3^{(k)}$ as well, and, in consequence, a 
further compensating shift in $\alpha_0$. From Eq.~(\ref{cfitproblem}),
the size of such effects, associated with the truncation of the running, 
and unavoidable at some level, can 
be minimized by taking $Q_0$ as large as possible (achieved by working 
with the observable with the highest intrinsic BLM scale) and keeping 
$t_k$ from becoming too large (achieved by restricting one's attention, 
if possible, to a subset of finer lattices)~\cite{qualifyingfootnote}.

\subsection{\label{subsec3}More on the relation between the two 
implementations}

For $n_f=3$, in our notation, the relation between $\alpha_V^p$ and 
$\alpha_s$, to $O(\alpha_s^3)$, is~\cite{schroder} 
\begin{equation}
\alpha_V^p(q^2)=\alpha_s(\mu^2)\left[1+\kappa_1(\mu^2/q^2)\alpha_s(\mu^2)
+\kappa_2(\mu^2/q^2)\alpha_s(\mu^2)\right]
\label{schrodereqn}\end{equation}
where $\kappa_2(x)=\left[ a_2 +16\beta_0^2\log^2(x)
+(16\beta_1+8\beta_0 a_1)\log (x)\right] /16\pi^2$,
with $a_2={\frac{695}{6}}\, +\, 36\pi^2\, -\, {\frac{9}{4}}\pi^4
+14\zeta (3)$, and $\kappa_1(x)=\left[ 7+4\beta_0\log (x)\right] /4\pi$.
Our expansion parameter, $\alpha_T(q^2)$ is defined to be equal to the
RHS of Eq.~(\ref{schrodereqn}) with $\mu^2 = q^2$,
leading to the numerical result given in Eq.~(\ref{alphatdefn}).
The conversion from $\alpha_T$ to $\alpha_s$
can be performed exactly but the absence in $\beta^T_{4,5,\cdots}$ 
of terms $\propto \beta_{4,5,\cdots}$
induces a perturbative uncertainty in the values
of our fitted parameters, one which can, however, 
be reduced by working with high scale observables
and fine lattices. It is also possible to test for its presence
by expanding the fits to include coarser lattices, where the effects
of the omitted contributions will be larger.

The construction of the expansion parameter $\alpha_V$ is somewhat more 
complicated, but turns out to be equivalent to the following~\cite{thanks}. 
One first takes the RHS of Eq.~(\ref{schrodereqn}),
with $\mu^2=e^{-5/3}q^2$, to define an intermediate
coupling, $\alpha_V^\prime (q^2)$. The corresponding $\beta$ function, 
$\beta^\prime$, is then determined to 4-loops by
$\beta_0,\cdots ,\beta_3$. The higher order coefficients,
$\beta^\prime_{4,5,\cdots}$, however, depend on the
presently unknown $\beta_{4,5,\cdots}$, are hence are themselves unknown.
The final HPQCD coupling, $\alpha_V$, is obtained from $\alpha_V^\prime$
by adding terms of $O(\alpha_s^5)$ and higher with coefficients chosen
to make $\beta^V_4=\beta^V_5=\cdots =0$. Since
$\beta_{4,5,\cdots}$ are not known, the values of the coefficients
needed to implement these constraints are also not known.
The coupling is nonetheless, in principle, well-defined, with
higher order coefficients computable as soon as
the corresponding higher order $\beta_k$ become available. Since
the 4-loop-truncated $\beta^V$ function is, by defnition, exact,
the distortions of the fit parameters induced, in general,
by the 4-loop truncation of the running are absent for the $\alpha_V$ coupling.
The price to be paid for this advantage is the unknown perturbative 
uncertainty in the relation between $\alpha_V$ and $\alpha_s$, which 
affects the conversion and running to $\alpha_s(M_Z)$. 
With this definition, $\alpha_V$ differs from $\alpha_T$ 
beginning at $O(\alpha_s^4)$.

The other difference between the two re-analyses lies in the treatment of 
$r_1/a$, $r_1$, and $\langle \alpha_s G^2/\pi \rangle$.
In Ref.~\cite{latticealphas}, these are allowed to vary independently, 
though within the range of the input prior constraints, for each $O_k$,
whereas in our analysis, they are treated as fixed external
input, and have the same central values for all $O_k$.
As noted above, the difference in the treatment of $r_1/a$ and $r_1$
is expected to have a negligible impact. The impact of the
differing treatments of $\langle \alpha_s G^2/\pi \rangle$
should be similarly negligible for observables with intrinsic scales 
high enough that the associated correction is small.

The two different implementations of the original HPQCD/UKQCD
approach will thus, when restricted to high-scale observables, 
correspond to isolating residual higher order
perturbative uncertainties in different sectors of the analysis.
If these uncertainties are, as desired, small in both cases, the 
two analyses should be in good agreement. Such agreement
(which is, in fact, observed, provided comparison is
made to the very recent HPQCD update) serves to increase
confidence in the results of both analyses.

\section{\label{sec3}Details of our re-analysis}

In our analysis, we have calculated the desired Wilson loops
using the publicly available $a\sim 0.09,\, 0.12,\,
0.15$ and $0.18$ fm MILC $n_f=2+1$ ensembles and incorporated
information on $W_{11}$ and $W_{12}$ for the three $a\sim 0.06$ fm 
USQCD ensembles provided to us by Doug Toussaint of the collaboration. 

We follow the basic strategy
of the earlier analysis, using the same 3-loop perturbative input, but
with the following differences in implementation. First, we
employ the expansion parameter $\alpha_T$ throughout. All running of 
$\alpha_T$ is carried out using exact 4-loop-truncated
running of the intermediate variable, $\alpha_s$, whose relation to
$\alpha_T$ is given by Eq.~(\ref{alphatdefn}). Second, to minimize
the effect of our incomplete knowledge of the running of $\alpha_T$
beyond 4-loop order, the impact of which will be larger for coarser lattices, 
we perform ``central'' 3-fold versions of our fits using the three finest
lattices, with $a\sim 0.12$, $0.09$ and $0.06$ fm. Expanded
5-fold fits then serve as a way of studying the impact
of the truncated running, as well as of the truncation
of the perturbative expansion for the $O_k$.
Since we do not currently have access to the actual $a\sim 0.06$ fm 
configurations, we are restricted to analyzing the three observables indicated
above. One of these, $\log(W_{12}/u_0^6)$, has a significantly
lower BLM scale, and hence is particularly useful for studying
the impact of these truncations.
As in Ref.~\cite{latticealphas}, we extrapolate linearly in the quark 
masses~\cite{linearextrapfootnote},
and estimate (and subtract) residual mass-independent NP effects
using the known form of the leading order gluon condensate contributions
to the relevant Wilson loops. 

Regarding the mass extrapolation, the sets of configurations for 
different mass combinations $am_\ell /am_s$ corresponding to approximately
the same lattice spacing $a$ in fact have slightly
different measured $r_1/a$. Since the $O_k$ we
study are themselves scale-dependent, full consistency requires
converting the results corresponding to the different $am_\ell /am_s$
to a common scale before extrapolation. This could be done with
high accuracy if the parameters appearing in the 
perturbative expansion of the $O_k$ were already known. Since, however,
some of these parameters are to be determined as part of
the fit, the extrapolation and fitting procedure must be iterated. 
With sensible starting points, convergence is achieved
in a few iterations. The dominant uncertainty in the converged iterated
extrapolated values is that associated with the uncertainties in 
$r_1/a$. There is also a $100\%$-correlated global scale uncertainty
associated with that on $r_1$. We employ $r_1=0.318(7)$ fm, as 
given in the MILC Lattice 2007 pseudoscalar project update~\cite{bernard07}.

The mass-independent NP subtractions are estimated using the leading 
order (LO) $D=4$ gluon condensate contribution, $\delta_g W_{mn}$, 
to the $m\times n$ Wilson loop, $W_{mn}$~\cite{wmnglue}
\begin{equation}
\delta_g W_{mn}\, =\, {\frac{-\pi^2}{36}}m^2n^2 a^4
\langle \alpha_s G^2/\pi\rangle
\label{gluewmnterm}\end{equation}
and the central value, $\langle \alpha_s G^2/\pi\rangle\, 
=\, (0.009\pm 0.007)\ {\rm GeV}^4$, 
of the updated charmonium sum rule 
analysis~\cite{newgcond4}.
Since the error here is already close to $100\%$,
we take the difference between results obtained
with and without the related subtraction as a measure of the 
associated uncertainty. This should be sufficiently conservative
if the correction is small. If not,
the measured $O_k$ values may contain additional non-negligible 
mass-independent contributions, of dimension $D>4$, which we do 
not know how to estimate and subtract. $O_k$
for which this occurs will thus provide a less reliable determination
of $\alpha_s$.

Fortunately, for the observables we consider, the gluon condensate
correction is, as desired, small. For $O_k=\log (W_{11})$, the corrections 
required for the 3-fold (5-fold) fit do not exceed $\sim 0.1\%$ ($\sim 0.5\%$).
The corrections remain small (less than $\sim 0.4\%$ ($\sim 1.8\%$)) for 
$O_k=\log (W_{12})$. The effect is somewhat larger for $\log (W_{12}/u_0^6)$,
as a consequence of cancellations encountered in combining the uncorrected
$\log (W_{11})$ and $\log (W_{12})$ values, but still reaches only
$\sim 1.3\%$ ($\sim 5.6\%$) for the 3-fold (5-fold) fit~\cite{gluefootnote}.

In line with what was seen in Ref.~\cite{latticealphas}, we find that the 
known terms in the perturbative expansions of the $O_k$ are
insufficient to provide a description of the observed scale-dependence,
even when only the three finest lattices are considered. 
When $c_3^{(k)}$ is added to the fit, however,
we find very good fits, with $\chi^2/dof<1$ (very significantly so
for the 3-fold fits). With current errors, it is thus
not possible to sensibly fit additional coefficients $c_{m>3}^{(k)}$.
This raises concerns about possible truncation uncertainties. 
Comparison of the results of the 3-fold and 5-fold fits
provides one handle on such an uncertainty since the relative weight of
higher order to lower order terms grows with decreasing
scale. If neglected higher order terms are in fact {\it not} negligible,
the growth with decreasing scale of the resulting fractional error 
should show up as an instability in the values of the parameters extracted 
using the different fits. We see no signs for such an instability within the
errors of our fits, but nonetheless include 
the difference of central values obtained from the 3-fold and 5-fold
fits as a component of our error estimate.

\section{\label{sec4}Results}
Central inputs for our fits are the measured
lattice observables (whose errors are tiny on the scale of
the other uncertainties), the computed $D_k$,
$c_1^{(k)}$ and $c_2^{(k)}$~\cite{latticealphas,mason}, 
$r_1/a$, $r_1$ and $\langle \alpha_s G^2/\pi \rangle$, 
and the choice of the 3-fold fitting procedure. In addition to
the uncertainties generated by the errors on 
$r_1/a$, $r_1$ and $\langle \alpha_s G^2/\pi \rangle$, 
are those due to uncertainties in numerical evaluations of the 
$D_k$, $c_1^{(k)}$ and $c_2^{(k)}$. 

We construct an ``overall scale uncertainty error'' by adding {\it linearly}
the fit uncertainties generated by those on $r_1$ and the $r_1/a$. This 
combined error is added in quadrature to (1) uncertainties produced by 
varying the $c_2^{(k)}$ (and, if relevant, $c_1^{(k)}$) within
their errors, (2) the difference between results obtained with and
without the gluon condensate correction, and (3) the difference between
the results of the 3-fold and 5-fold fits. Because of the
iterative nature of the fit procedure, the mass
extrapolation uncertainty is incorporated into what we
have here identified as the overall scale uncertainty. 

We run our
$n_f=3$ results to $M_Z$ using the self-consistent combination
of 4-loop running and 3-loop matching at the flavor thresholds, taking
the flavor thresholds to lie at $rm_c(m_c)$ and $rm_b(m_b)$, with
$m_c(m_c)=1.286(13)$ GeV and $m_b(m_b)=4.164(25)$ GeV~\cite{kss07},
and $r$ allowed to vary between $1$ and $3$. These uncertainties
in the matching thresholds, together with standard estimates for the 
impact of the truncated running and matching, produce 
an evolution contribution to the uncertainty on $\alpha_s(M_Z)$
of $\pm 0.0003$~\cite{bck08}.

Our central fit results for $\alpha_s(M_Z)$ and the $c_3^{(k)}$ are
given in Table~\ref{resultstable}. For comparison, the results for
$\alpha_s(M_Z)$ obtained in Ref.~\cite{latticealphas} were 
$0.1171(12)$, $0.1170(12)$ and $0.1162(12)$, for 
$\log(W_{11})$, $\log(W_{12})$ and $\log(W_{12}/u_0^6)$, respectively. Our 
$\alpha_s(M_Z)$ are significantly larger, and in closer mutual agreement. The 
recent HPQCD update~\cite{hpqcd08} also finds significantly larger values. 
(We will return to a more detailed comparison of the two updates below.)
The very good agreement between the $\alpha_s(M_Z)$ values obtained in our 
fits using both low- and high-scale observables suggests 
that the effects of the truncated running, present at some
level in all such fits, are small in the cases we have studied.

\begin{table}
\caption{\label{resultstable}Central fit results for 
$\alpha_s(M_Z)$ and the $c_3^{(k)}$}
\vskip .1in
\begin{tabular}{|c|c|c|}
\hline
$O_k$&$\alpha_s(M_Z)$&$c_3^{(k)}$\\
\hline
$\log\left( W_{11}\right)$&$\ \ 0.1192\pm 0.0011\ \ $&$-3.8\pm 0.6$\\
$\log\left( W_{12}\right)$&$\ \ 0.1193\pm 0.0011\ \ $&$-4.0\pm 0.9$\\
$\log\left( W_{12}/u_0^6\right)$&$\ \ 0.1193\pm 0.0011\ \ $&
$-1.7\pm 0.8$\\
\hline
\end{tabular}
\end{table}

\begin{figure*}
  \begin{minipage}[t]{0.48\linewidth}
\includegraphics[width=0.9\textwidth]{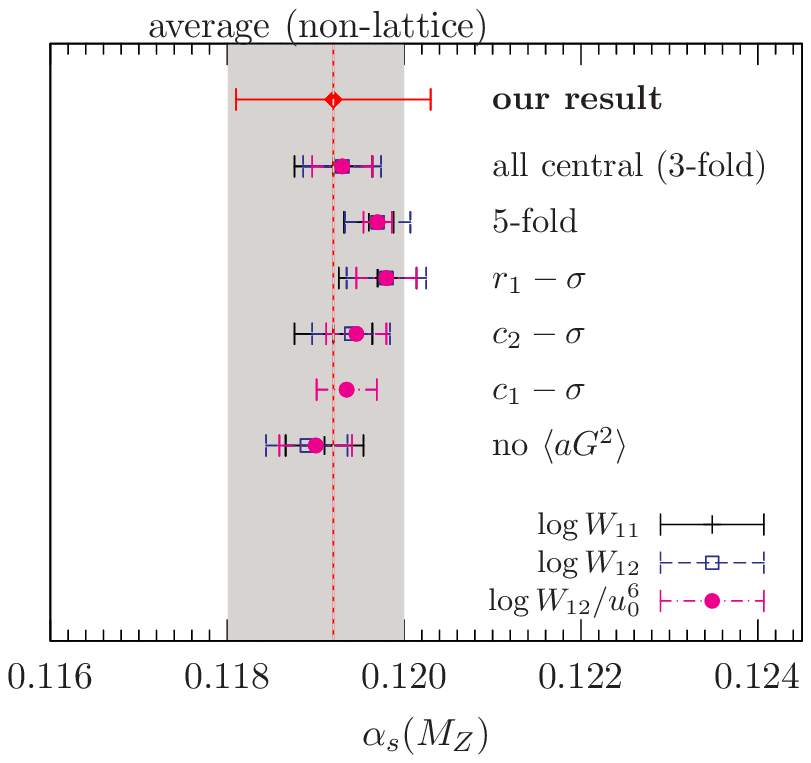}
  \caption{Contributions to the errors on $\alpha_s(M_Z)$. Shown are
          the $\alpha_s(M_Z)$ obtained using (i) the 3-fold
          fit strategy, with all central input, 
          (ii) the alternate 5-fold fit strategy,
          with all central input, and (iii) the 3-fold
          fit strategy, with, one at a time, each input
          shifted from its central value by $1\sigma$, retaining central
          values for the remaining input parameters. The error bars shown
          are those associated with the uncertainties in $r_1/a$.}
  \label{figure1}
  \end{minipage}
\hfill
  \begin{minipage}[t]{0.48\linewidth}
\includegraphics[width=0.9\textwidth]{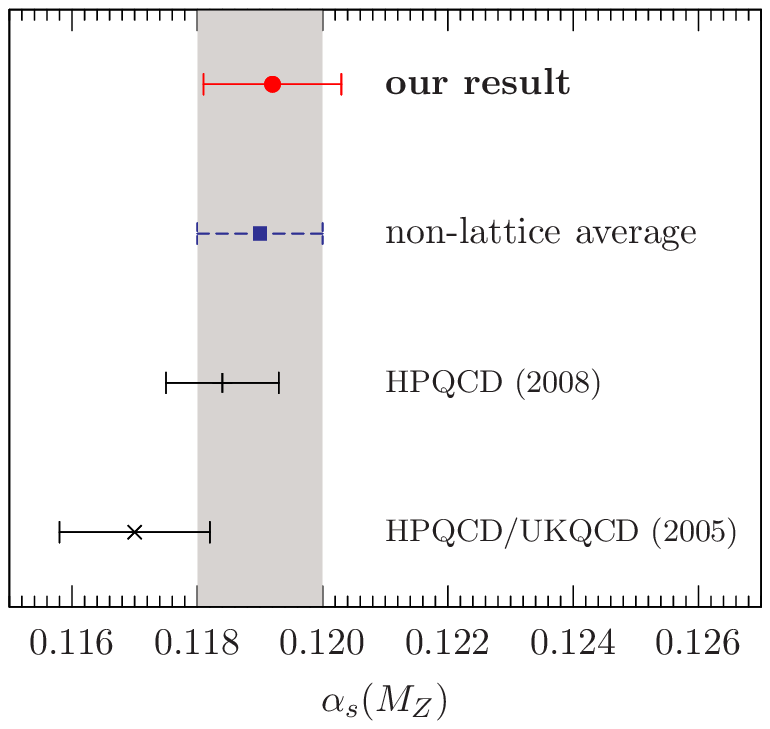}
   \caption{Comparison of the results for $\alpha_s(M_Z)$ from our fits,
            the fits of Ref.~\cite{latticealphas} and the updated fits
            of Ref.~\cite{hpqcd08} with the average of recent non-lattice
            determinations.}
   \label{figure2}
  \end{minipage}
\end{figure*}

One-sided versions of
the various components of the total errors on $\alpha_s(M_Z)$ 
are displayed in 
Figure~\ref{figure1}. 
The difference of the 3-fold and
5-fold determinations is $\sim 0.0004$, 
significantly smaller than the $\sim 0.0009$ overall scale
uncertainty. The results thus show no evidence for any instability
associated with opening up the fit to lower scales.

While the total error on $\alpha_s(M_Z)$ 
is the same for all three $O_k$ considered,
the general arguments above lead us to believe that the
most reliable determination is that obtained using the
highest-scale observable, $\log(W_{11})$, and highest-scale 
(3-fold fit) analysis window. Our
final assessment,
\begin{equation}
\alpha_s(M_Z)=0.1192\pm 0.0011\ ,
\end{equation}
is in excellent agreement with the non-lattice average and the
result, $0.1184\pm 0.0009$, of the independent
HPQCD analysis. The various results are shown for
comparison in Figure~\ref{figure2}. A more detailed discussion
of the relation between our re-analysis and that of HPQCD may
be found in the Appendix.

\begin{acknowledgments}
This work was supported by the Natural Sciences and Engineering Research
Council of Canada, the Australian Research Council, eResearch
South Australia, and the Australian Partnership for Advanced Computing. 
Thanks to the members of the MILC Collaboration for making
their configurations available to the community, and to Doug Toussaint and
Carleton de Tar for providing information on the $a\sim 0.06$ and
$a\sim 0.15$ fm lattices not currently available in the literature.
KM would like to acknowledge very useful discussions with the
members of the HPQCD collaboration, especially G.P. Lepage, clarifying
details of the original lattice analysis, the recent HPQCD
update, and providing information on ongoing work.

\end{acknowledgments}

\appendix*
\section{More on the relation to the HPQCD re-analysis}

After the completion of the work reported in this
paper, the HPQCD Collaboration posted an update of their earlier
2005 analysis~\cite{hpqcd08}. This update works with a subset of 11 of the
available MILC ensembles, spanning the $a\sim 0.18$, $0.15$, $0.12$,
$0.09$ and $0.06$ fm lattices and a range of $am_\ell /am_s$.
The fits follow the strategy of the earlier analysis~\cite{latticealphas},
employing the expansion parameter $\alpha_V$, and fitting the
unknown $\bar{c}^{(k)}_n$ using priors. Linear mass extrapolation has been
employed, and mass-independent NP $D=4$ contributions 
estimated and subtracted using the LO formula for $\delta_g W_{mn}$.
The fitting of $r_1/a$, $r_1$ and $\langle \alpha_s G^2/\pi\rangle$,
observable by observable, using central input and prior widths, 
is also as in the earlier analysis,
with the exception that the central value and width for 
$\langle \alpha_s G^2/\pi\rangle$ are now
$0$ and $\pm 0.012\ {\rm GeV}^4$, respectively.

The HPQCD implementation differs from ours in the
choice of expansion parameter, and in the implementation of the input 
information on $r_1/a$, $r_1$ and $\langle {\frac{\alpha_s}{\pi}}G^2\rangle$.
For the reasons discussed above, we expect the impact on $\alpha_s(M_Z)$ 
of the observable-by-observable fitting of $r_1/a$, $r_1$ and 
$\langle \alpha_s G^2/\pi \rangle$ in the HPQCD approach to be small
for $O_k$ having small gluon condensate corrections. Since the 
different choices of expansion parameter correspond to different ways 
of isolating residual higher-order perturbative uncertainties,
one expects the results of the two analyses to be in good agreement 
so long as (i) one is working with $O_k$ having small mass-independent 
NP corrections, (ii) the same input values are used for both, 
and (iii) residual NP and higher-order perturbative
uncertainties are indeed small. The situation is likely to be more
complicated for $O_k$ with sizeable estimated $D=4$ gluon condensate 
corrections.

The results of the HPQCD fit for the three $O_k$ we consider
are $\alpha_s(M_Z)=0.1186(9)$,
$0.1186(9)$ and $0.1183(8)$ for $\log (W_{11})$, $\log (W_{12})$
and $\log (W_{12}/u_0^6)$, respectively~\cite{hpqcd08}. All are in good 
agreement within errors with the corresponding results from our analysis.
This agreement is further improved if one
takes into account the small difference in input $r_1$ values.
Were we to switch from $r_1=0.318$ fm to the central value of 
the HPQCD determination, $0.321(5)$ fm, all three of our 
$\alpha_s(M_Z)$ results would decrease by $0.0002$. Note also that use of 
the central charmonium sum rule input for $\langle \alpha_s G^2/\pi\rangle$ 
in our calculation raises the output $\alpha_s(M_Z)$ obtained from 
$\log (W_{11})$, $\log (W_{12})$ and $\log (W_{12}/u_0^6)$ by $0.0001$, 
$0.0004$ and $0.0005$, respectively. Our fitted values
would thus be in even closer agreement with those of the
HPQCD update were we to impose the HPQCD central zero value 
of $\langle \alpha_s G^2/\pi\rangle$ in our fits.
Since the fitted $\langle \alpha_s G^2/\pi \rangle$ values obtained by
analyzing the various $O_k$ are not quoted in Ref.~\cite{hpqcd08}, 
it is not possible to quantify further the role of this effect. The 
agreement for the three observables under discussion is in any case good, 
within expectations, independent of this question.

We now turn to a more detailed discussion of the issue of the
subtraction of the mass-independent NP contributions.
If the estimated LO, $D=4$ gluon condensate subtraction
represents only a small fraction of the measured $O_k$ at
the scales under consideration, analogous mass-independent NP 
contributions with $D>4$ should be even smaller, and hence 
safely negligible. If, however, the estimated $D=4$ correction
is sizeable, analogous $D>4$ corrections can no longer be expected
to be small. These necessarily scale differently with 
lattice spacing than do the $D=0$ perturbative and $D=4$
NP contributions and hence, if not included when fitting the data,
are likely to force shifts in both $\alpha_s$ and 
$\langle \alpha_s G^2/\pi \rangle$ if present at a non-negligible level. 

We deal with this potential problem by focussing on $O_k$
for which the impact of the estimated $D=4$ gluon condensate subtraction
is small compared to the variation of the $O_k$ in question over the 
lattice scales employed in the fit. In the initial version of the 
HPQCD re-analysis, mass-independent NP subtractions
were estimated using only the $D=4$ gluon condensate form, even
for observables where the estimated correction is sizeable.
In the more recent update, additional terms, scaling as would 
mass-independent contributions of $D>4$, are added to the fit function 
for each observable, and the accompanying coefficients extracted as part 
of the augmented Bayesian fit. The impact of including the 
$D>4$ terms is, as expected, small for those observables having small values 
of the estimated $D=4$ subtraction. For observables with larger
$D=4$ subtractions, the fit errors are increased (by factors of
$\sim 2$ for those observables having the largest $D=4$ corrections)
and some shifts in $\alpha_s(M_Z)$ of order $1/2$ to $1$ times the 
smaller preliminary errors are observed. The shifts serve to reduce
the spread of $\alpha_s(M_Z)$ values compared to that seen in
the original version of the re-analysis. The values of
$\langle \alpha_s G^2/\pi\rangle$ obtained from the independent fits to the
different observables are not quoted in Ref.~\cite{hpqcd08}, but
a useful test of the self-consistency of the approach
would be to verify that the inclusion of the $D>4$
contributions has brought these values into good agreement with one another.

It is worth noting that
the observables, $\log (W_{23}/u_0^{10})$, $\log (W_{14}/W_{23})$,
and $\log (W_{11}W_{23}/W_{12}W_{13})$, which produce the three
smallest results for $\alpha_s(M_Z)$, have estimated $D=4$ corrections 
significantly larger than those for any of the other observables.
The magnitudes of the corrections in these cases represent 
$\sim 50-100\%$ of the variation with scale of the uncorrected
$O_k$ between the lightest mass $a\sim 0.06$ and $a\sim 0.12$ fm
ensembles. (This variation-with-scale provides a suitable measure for use
in assessing the importance of NP corrections 
since it is the variation with scale which provides the input needed to
fix the fit parameters, and, as explained in Ref.~\cite{hpqcd08}, the 
$a\sim 0.06,\, 0.09$ and $0.12$ fm ensembles which dominate the HPQCD 
re-analysis.) These observables are thus, for the purposes of the analysis,
rather non-perturbative. Were one to exclude observables
with larger NP contributions from the HPQCD average, on
the grounds that the related subtractions introduce additional
theoretical systematic uncertainties, the HPQCD result would
be brought into even closer agreement with ours, though the
resulting shift would in fact be small (at the $\sim$ quarter $\sigma$ level).

We stress that, independent of these questions, our results agree well 
within errors with those of the HPQCD update. This agreement is further
improved by a shift to common input. We argue 
that the non-zero central value for $\langle \alpha_s G^2/\pi \rangle$
obtained from the updated charmonium sum rule analysis
represents our best present knowledge of this quantity, and hence
also the best choice as input for evaluating the small
mass-independent NP subtractions needed
for extracting $\alpha_s(M_Z)$. In addition, for the reasons
just discussed, we believe that the most reliable
determinations of $\alpha_s(M_Z)$ are those
based on those observables for which the $D=4$
correction is as small as possible. Such an assessment
produces the results already noted above, which are in
extremely good agreement with what it known from other sources.


\vfill\eject


\begin{thebibliography}{99}
\bibitem{tauentryfootnote}We have quoted, to be specific, the
$\tau$ decay result of Ref.~\cite{my08}, which contains a
careful treatment of the small, but (for a precision
determination) numerically important, $D>4$ OPE contributions,
the inconsistent treatment of which 
created unquantified theoretical systematic errors in a 
number of the earlier $\tau$-decay-based analyses. The results of the fifth 
order FOPT analysis, $0.1185^{+0.0014}_{-0.0009}$, and of the
more model dependent extension, $0.1180\pm 0.0008$,
both reported in Ref.~\cite{jb08}, are compatible within errors
with that of Ref.~\cite{my08}, but are based on assumed, rather than
fitted, values for the relevant $D=6$ and $8$ OPE contributions.
The ranges corresponding to the two results of Ref.~\cite{jb08},
however, are fully contained within the range listed in the table.
\bibitem{aggthanks}The details on the treatment of 
$\langle \alpha_s G^2/\pi\rangle$ reported here are not contained 
in Ref.~\cite{latticealphas}. Thanks to G.P. Legage of the HPQCD 
collaboration for providing us with this information.
\bibitem{matchingrunningfootnote}To be explict, a
self-consistent combination of 2-loop matching and 3-loop 
running~\cite{cks97} was employed, with the $n_f=3\rightarrow 4$ and 
$n_f=4\rightarrow 5$ flavor thresholds taken to be at $m_c(m_c)=1.25$ GeV 
and $m_b(m_b)=4.25$ GeV, respectively. The use of 2-loop matching and
3-loop running rather than 3-loop matching and 4-loop running
is inessential; the resulting value at scale $M_Z$ is
lowered by only $0.0002$ if one uses the latter approach,
a change which is small on the scale of the final quoted
error.
\bibitem{qualifyingfootnote}The non-universal $\beta^T_{n>1}$ 
receive contributions from all of the $\overline{MS}$
coefficients $\beta_m$ with $m\leq n$. Numerically, one has
$\beta^T_2=\beta_2+23.9090=33.9689$, $\beta^T_3=\beta_3-371.6215=-324.3934$,
$\beta^T_4=\beta_4-758.1717$, $\beta^T_5=\beta_5-7\beta_4+30807.8757$,
$\cdots$, where the additive numerical terms reflect the
known contributions of the lower order $\beta_k$. Since
terms involving $\beta_0,\cdots ,\beta_3$ are kept to all orders
in our analysis, only the parts of $\beta^T_4,\beta^T_5,\cdots$ 
proportional to $\beta_4,\beta_5,\cdots$ are not properly included.
The expressions for $\beta^T_{2,3}$, show that the lower
order terms in fact dominate in those cases. The expressions
for $\beta^T_{4,5}$ will also be dominated by the already-fully-included
lower order contributions unless an extremely rapid growth with order 
occurs for the $\overline{MS}$ $\beta$ function coefficients. The 
distortion of the $t_k$-dependence discussed in the text should thus, 
in fact, be considerably weaker than it might appear from the original 
form of the argument. It nonetheless 
remains sensible to treat with greater confidence analyses which are
based on high scale observables and employ the finest lattices.
\bibitem{gluefootnote}The relative size of the gluon condensate correction 
grows rapidly with loop dimension. 
For the $2\times 3$ loop, for example, the correction to $O_k=\log (W_{23})$, 
would reach $\sim 8\%$ for the 3-fold fit, and $\sim 31\%$ for the coarsest 
of the lattices entering the 5-fold fit (where the ``correction'', 
$\delta_g W_{23}$, has, in fact, grown to a factor of $\sim 2$ larger in 
size than the measured $W_{23}$ itself). A further enhancement,
similar to that encountered for $\log (W_{12}/u_0^6)$, occurs
when $O_k$ is the logarithm of one of the Creutz ratios or tadpole-improved 
ratios. For example, the size of the ``corrections'' for
$O_k = \log (W_{23}/u_0^{10})$, $\log (W_{11}W_{23}/W_{12}W_{13})$ 
and $\log (W_{14}/W_{23})$, which represent the three extreme ``outliers'' 
of the original HPQCD/UKQCD analysis, are $56\%$, $82\%$ and 
$101\%$, respectively.
\end{thebibliography}
\end{document}